\documentclass[amssymb,prl,aps,twocolumn,amsmath]{revtex4} 

\usepackage{graphicx}

\begin{document}

\title{Dephasing by extremely dilute magnetic impurities revealed by Aharonov-Bohm oscillations}
\author{F. Pierre and Norman O. Birge}
\address{Department of Physics and Astronomy, Michigan State University, East Lansing, Michigan 48824-2320}
\date{\today}

\begin{abstract}
We have probed the magnetic field dependence of the electron phase
coherence time $\tau_\phi$ by measuring the Aharonov-Bohm
conductance oscillations of mesoscopic Cu rings. Whereas
$\tau_\phi$ determined from the low-field magnetoresistance
saturates below 1~K, the amplitude of Aharonov-Bohm $h/e$
oscillations increases strongly on a magnetic field scale
proportional to the temperature. This provides strong evidence
that a likely explanation for the frequently observed saturation
of $\tau_\phi$ at low temperature in weakly disordered metallic
thin films is the presence of extremely dilute magnetic
impurities.
\end{abstract}


\maketitle

Understanding the sources of decoherence is of fundamental
importance in all quantum systems that interact with their
environment. In mesoscopic physics, quantum phase coherence of
conduction electrons in disordered metals and semiconductors gives
rise to a wide variety of novel transport phenomena observed at
low temperature \cite{meso}; decoherence sets the limit on the
lengthscale over which such effects are observable. In metallic
thin films below about 1 K, the dominant source of decoherence is
thought to be electron-electron interactions. For narrow wires,
the phase coherence time for that process was calculated by
Altshuler, Aronov, and Khmelnitskii~\cite{AAK} in 1982, and
diverges at low temperature as $\tau_\phi \propto T^{-2/3}$.

This theoretical picture was deeply shaken in 1997 when two
different experiments suggested that electrons in mesoscopic
metallic wires interact with each other much more strongly than
predicted by theory. One experiment \cite{MJW} showed that
$\tau_\phi$ determined from the low field magnetoresistance of
mesoscopic Au wires, systematically saturates at low temperature,
rather than continuing to increase as a power law. The other
experiment \cite{PRLrelax} showed that the energy exchange rate
between electrons measured in several Cu wires exceeded the
prediction for electron-electron interactions \cite{AAK} and
obeyed a different energy dependence.

More recently, it was found that the measured energy exchange rate
and dephasing rate are sample dependent and tightly correlated to
the source material purity: samples fabricated using a very pure
(99.9999\%) Ag or Au source agreed with the theoretical prediction
for electron-electron interactions, whereas samples fabricated
with a source of lesser purity, or with Cu, showed a smaller phase
coherence time and a larger energy exchange rate
\cite{PierreAg,Gougam,PierrePesc,PierrePHD,PierrePRBWL}.

There have been several theoretical suggestions regarding the
source of the excess dephasing. Several of those suggestions, such
as the possibility of electromagnetic interference coming from
outside the cryostat \cite{AAG}, can be ruled out purely on
experimental grounds, since different samples measured in the same
cryostat show different behaviors \cite{Gougam}. The controversial
theory by Golubev, Zaikin, and Sch\"{o}n of zero temperature
dephasing by high energy electromagnetic modes \cite{Zaikin} is
able to account for only a subset of the experimental results
published in references \cite{MJW,Gougam,PierrePesc,Natelson},
using the overall prefactor of the dephasing rate as an adjustable
parameter. In our opinion, the only two explanations that have not
been ruled out by experiment are dephasing by two-level systems
\cite{IFS} and dephasing by paramagnetic impurities
\cite{dephasingMI}. We aim to demonstrate in this letter that all
the data showing $\tau_\phi$ saturation in weakly disordered
metallic thin films published to date might be explainable by the
presence of paramagnetic impurities at extremely low concentration
\cite{JJLin}.

How can we determine whether a sample which exhibits a saturation
of $\tau_\phi$ indeed contains unexpected magnetic impurities and
that $\tau_\phi$ is really limited by spin-flip scattering from
these extrinsic degrees of freedom?
\begin{table}[tbhp]
\begin{tabular}{|c|c|c|c|c|c|c|}
\hline Sample&$t$ (nm)&$w$ (nm)&$R$ ($\Omega$)&$D$
($\mathrm{cm}^{2}/\mathrm{s}$)&$r$ ($\mu \mathrm{m}$)& $w_{AB}$
(nm)\\\hline
Cu1 & 45 & 155 & 700 & 146 & n.a. & n.a.\\
Cu2 & 20 & 70 & 7980 & 61 & n.a. & n.a.\\
Cu3 & 33 & 75 & 4370 & 65 & 0.5 & 67\\
Cu4 & 20 & 80 & 8500 & 52 & 0.75 & 73\\
\hline
\end{tabular}
\caption{Geometrical and electrical characteristics of the
measured copper samples. The diffusion constant $D$ is obtained
using Einstein's relation with the density of states in copper
$\nu_F=1.56~10^{47}$ $\mathrm{J^{-1}m^{-3}}$ and the resistivity
$\rho$ extracted from the thickness $t$, length $L=285~\mu
\mathrm{m}$, width $w$ and resistance $R$ of the long wire. In
samples Cu3 and Cu4, the ring's radius and linewidth are
respectively $r$ and $w_{AB}$.} \label{table1}
\end{table}

\begin{figure}[htbp]
\includegraphics[width=3in]{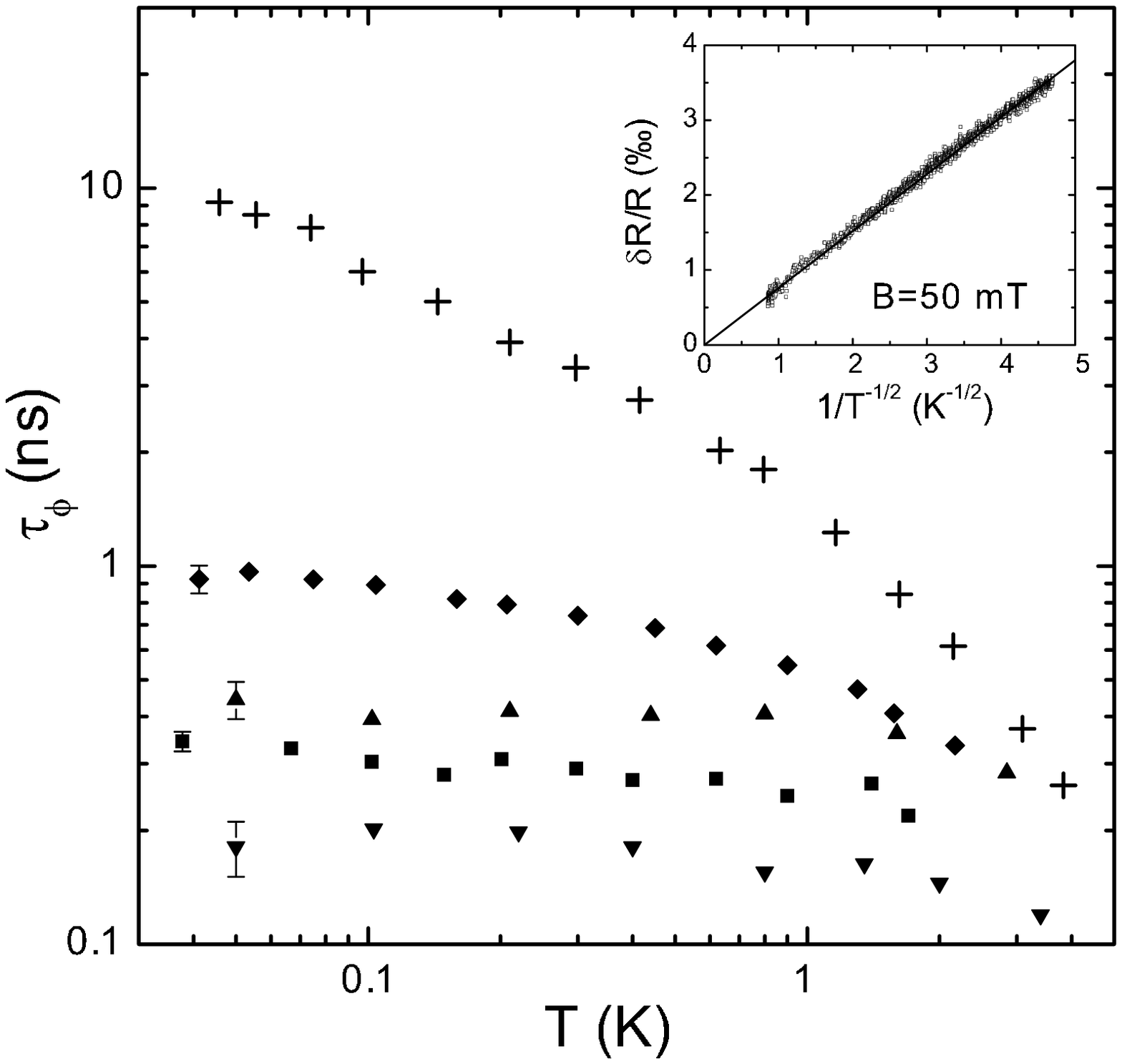}
\caption{Phase coherence time $\tau_\phi(B\approx 0)$ vs
temperature determined by fitting the low field magnetoresistance
of long wires to weak localization theory. The uncertainties are
indicated by error bars at the lowest temperature where they are
usually the largest. The samples are made of silver Ag1 ($+$)
(data taken from~[6]) and copper Cu1 ($\blacktriangle$), Cu2
($\blacklozenge$), Cu3 ($\blacktriangledown$) and Cu4
($\blacksquare$). $\tau_\phi(B\approx 0)$ increases continuously
with decreasing temperatures in the silver sample, whereas in all
Cu samples $\tau_\phi(B\approx 0)$ shows a saturation at low
temperatures. Inset: open symbols represent the relative variation
of the resistance of sample Cu4 in a small magnetic field
$B=50~\mathrm{mT},$ plotted as a function of $1/\sqrt{T}.$ The
continuous line is a fit using the functional form $A/\sqrt{T}$
predicted by the theory of electron-electron interaction in
diffusive metallic wires [17]. The best fit is obtained with
$A=7.6~10^{-4}~\mathrm{K}^{1/2}$ in close agreement with the
theoretical prediction $7.2~10^{-4}~\mathrm{K}^{1/2}.$}
\label{Fig1}
\end{figure}

Probing the presence of unknown magnetic impurities at a
concentration of 1 part per million (ppm) or smaller is a
difficult challenge. For instance, the logarithmic contribution to
the resistance by Kondo impurities \cite{Kondo} at such small
concentrations can not be reliably detected through the
measurement of $R(T)$: it is very small and often hidden by the
$1/\sqrt{T}$ contribution due to electron-electron interaction
\cite{AAK,AleinerWav}. Since the dephasing rate is so sensitive to
magnetic impurities, it is natural to use the dephasing rate
itself to detect extremely small amounts of such impurities. In
the presence of a sufficiently large magnetic field spin-flip
collisions are frozen out, hence $\tau_\phi$ should return to the
value expected from electron-electron interaction. One method is
to measure the weak-localization contribution to the perpendicular
magnetoresistance in the presence of a parallel field large enough
to freeze the spins. However this requires an accurate alignment
of the parallel field with the wire's axis, and the residual
magnetic flux through the wire's cross section complicates the
data analysis \cite{Giordano}. Another possibility is to measure
the universal conductance fluctuations (UCF) of a metallic wire as
a function of magnetic field $B$. But the weak dependence on
$\tau_\phi$ and broad width in $B$ of UCF are not well suited to
probe the presence of very dilute magnetic impurities at low
temperature. We have chosen a third method. We fabricate a
ring-shaped sample with connecting leads, and measure the
Aharonov-Bohm (AB) oscillations in the magnetoconductance of the
ring. Such measurements were pioneered by Webb and coworkers in
the 1980's, and observed by several other groups at about the same
time \cite{WebbAB}. To obtain the largest possible AB
oscillations, the rings measured at this time had a very small
diameter thereby reducing their sensitivity to the value of
$\tau_\phi$. Hence, it is not surprising that the AB oscillations
were found independent of the magnetic field except on a few
samples purposely contaminated with a relatively very large ($\sim
100$~ppm of Mn in Au) concentration of magnetic impurities
\cite{WebbABMI}.

We applied this technique, with carefully designed rings, to probe
the magnetic field dependence of $\tau_\phi$ in two Cu samples,
labelled Cu3 and Cu4 (see Table~1). The choice of copper followed
from three observations.  First, as illustrated in Fig.~1, all of
the samples we have made with this material show a saturation of
$\tau_\phi(B\approx0)$ at low temperature, regardless of the
source purity (see also \cite{Gougam,PierrePRBWL}). Second,
whereas it is possible to attain an apparent saturation of
$\tau_\phi$ over a limited temperature range from a combination of
electron-electron interaction and spin-flip scattering
\cite{PierrePRBWL}, the flat and small $\tau_\phi$ in samples Cu3
and Cu4 is difficult to reproduce in this manner. Third, no trace
of Kondo magnetic impurities could be detected in the temperature
dependence of the resistance, shown for sample Cu4 in the inset of
Fig.~1. These three observations cast doubt on our proposal that
the saturation of $\tau_\phi$ at low temperature in Cu could be
explained by magnetic impurities, hence Cu is an ideal candidate
for this experiment.

\begin{figure}[htbp]
\begin{center}
\includegraphics[width=2in]{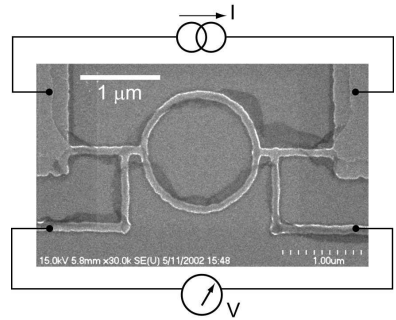}
\end{center}
\caption{Photograph of the Aharonov-Bohm ring in sample Cu4. We
obtain the resistance by a four-lead measurement. The bottom left
voltage lead is a very long wire used to extract
$\tau_\phi(B\approx0)$ from the low field magnetoresistance.}
\label{Fig2}
\end{figure}

We measured the magnetoconductance of two rings of radius $r=0.5$
and $r=0.75$~$\mu$m, located respectively on samples Cu3 and Cu4.
A scanning electron microscope picture of the ring in sample Cu4
is shown in Fig.~2. In addition to the ring, the samples Cu3 and
Cu4 contain a long meander line used to extract
$\tau_\phi(B\approx 0)$ by fitting the low-field magnetoresistance
with the prediction of weak localization theory \cite{AleinerWav}.
The fit procedure is detailed in \cite{fitWL}. All the Cu samples
were deposited on a silicon substrate, through a suspended mask
fabricated using standard e-beam lithography, in a thermal
evaporator used only for nonmagnetic metals Ag, Al, Au, Cu, Ti.
The evaporation rate ranges between 0.2 and
0.5~$\mathrm{nm}/\mathrm{s}$, under a pressure that stays below
$10^{-6}$~mbar. The source material and boat were replaced before
each evaporation and manipulated using plastic tweezers. Before
the actual deposition on the Si substrate we melt the Cu source,
evaporate 10-20~nm onto a shutter, and pump the chamber down again
for about 15~mn. This pre-evaporation covers the walls of the
evaporator with a clean layer of Cu and makes it possible to
maintain a very low pressure during the sample fabrication. The
samples were immersed in the mixing chamber of a dilution
refrigerator and measured through filtered electrical lines.
Resistance measurements were performed using a standard ac
four-terminal technique with a lock-in amplifier. The ac voltage
excitation $V_{ac}$ across the sample satisfies $eV_{ac} \lesssim
k_BT$ to avoid heating of electrons.

\begin{figure}[tbhp]
\includegraphics[width=3in]{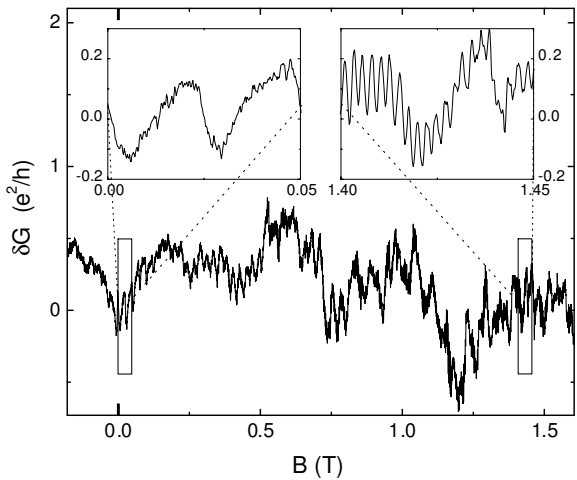}
\caption{Measured conductance of the ring in sample Cu4, in units
of $e^2/h$, as a function of magnetic field at a temperature
$T=100$~mK. The narrow Aharonov-Bohm oscillations ($\Delta
B\simeq2.5$~mT) are superimposed on the larger and much broader
universal conductance fluctuations. Left inset: blowup of the data
near zero field. The AB oscillations are hardly visible. Right
inset: blowup of the data at large magnetic field. The AB
oscillations are much larger.} \label{Fig3}
\end{figure}

The most important result of this article is visually striking in
the raw magnetoconductance data, as illustrated on Fig.~3 for
sample Cu4 at $T=100$~mK. Whereas the Aharonov-Bohm oscillations
can hardly be seen at small magnetic field $B$ (left inset in
Fig.~3), they are obvious at large $B$ (right inset in Fig.~3).
For a given geometry and temperature, the amplitude of AB
oscillations depends only on $\tau_\phi$, which means, without
further analysis, that $\tau_\phi$ increases with magnetic field
in this sample. A similar behavior was observed in sample Cu3.

To put this qualitative observation on solid ground we analyzed
the data by taking the Fourier transforms of magnetoconductance
data segments of width 0.2~T, using a Welch window. We define the
amplitude of the AB $h/e$ oscillations $\Delta G_{h/e}$ as the
square root of the power integrated over a narrow interval
surrounding the frequency $1/\Delta B\simeq \pi r^2/(h/e)$. Figure
4 shows the results for sample Cu4 at $T=40$ and 100~mK, as a
function of the reduced magnetic field $2\mu_BB/k_BT$. Apart from
confirming quantitatively the increase of AB oscillations with the
magnetic field, Fig.~4 shows that the magnetic field scale on
which the AB $h/e$ oscillations increase is proportional to the
temperature. This proves that the increase of AB oscillations at
large field results from the freezing of paramagnetic degrees of
freedom in the sample.

\begin{figure}[htbp]
\includegraphics[width=3in]{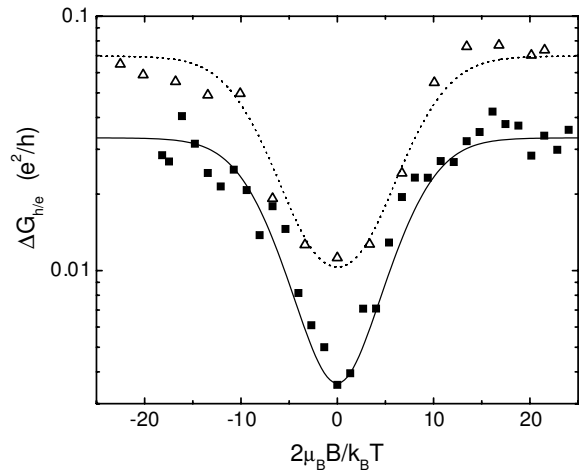}
\caption{Symbols: mean amplitude of the AB $h/e$ oscillations
($\Delta G_{h/e}$) in units of $e^2/h$ in sample Cu4 at $T=40$
($\vartriangle$) and 100~mK ($\blacksquare$), plotted as a
function of the reduced magnetic field $2\mu_BB/k_BT.$ Lines: fits
to the two data sets using Eqs.~(1), (2) and (3) with $C$ and $g$
as fit parameters (see text).} \label{Fig4}
\end{figure}

We can compare the measured amplitude of AB $h/e$ oscillations
with the prediction \cite{NoteAB}
\begin{equation} \label{eqGhe}
\Delta G_{h/e} = C \frac{e^2}{h} \frac{L_T}{\pi
r}\sqrt{\frac{L_\phi}{\pi r}}\exp{[-\pi r /L_{\phi}]},
\end{equation}
where $L_T=\sqrt{\hbar D/k_B T}$ is the thermal length, $D$ is
the diffusion coefficient, $T$ is the temperature, $L_\phi=\sqrt{D
\tau_\phi}$, $r$ is the ring radius and $C$ is a factor of order 1
that depends on the sample geometry in the vicinity of the ring.
The total dephasing rate $1/\tau_\phi$ is the sum of contributions
from electron-electron \cite{AleinerWav} and spin-flip scattering:

\begin{equation} \label{eqTauFi}
\frac{1}{\tau_\phi} = \frac{1}{\tau_{ee}}+\frac{1}{\tau_{sf}}.
\end{equation}
Whereas $1/\tau_{ee}$ is independent of $B$, the spin-flip
scattering rate vanishes at large field as \cite{Falko}
\begin{equation} \label{eqTauB}
\frac{\tau_{sf}(B=0)}{\tau_{sf}(B)} = \frac{g \mu_B B/k_B
T}{\sinh(g \mu_B B/k_B T)},
\end{equation}
where $g$ is the renormalized gyromagnetic factor of the magnetic
impurities.

The continuous line in Fig.~4 is a fit of $\Delta G_{h/e}(B)$ at
$T=100$~mK using Eqs.~(1), (2) and (3) with the gyromagnetic
factor $g=1.05$ and the overall multiplicative constant $C=0.13$
as the only fit parameters. The dotted line is the calculated
$\Delta G_{h/e}$ at $T=40$~mK using the same values for $g$ and
$C$, and taking into account the non negligible noise level of the
measurement at this temperature. We used the predicted
contribution for electron-electron interactions \cite{AleinerWav}:
$\tau_{ee}=7.4$~ns and 5.4~ns, respectively at $T=40$ and 100~mK,
and we used the values $\tau_{sf}(B=0) \simeq
\tau_\phi(B\approx0)$ \cite{NoteTausf} obtained from the low field
magnetoresistance of the long wire.

What can be these paramagnetic ``impurities" in our samples? It is
important to emphasize that the sample purity is not automatically
identical to the source material purity. Indeed, the residual
pressure in the evaporator is not small enough to rule out extra
contamination at the level of one ppm during evaporation. The
nature of such contaminants depends on the evaporator history.
However, our best candidate is the surface oxide of Cu
\cite{Shirane}. It was already pointed out in the late 1980's that
the surface of Cu may cause dephasing \cite{Haesendonck}. By
comparing the value of $\tau_\phi(B \approx 0)$ at low temperature
with the unitary limit ($T=T_K$) of spin-flip scattering in the
Kondo regime \cite{NagSuhl,PierrePRBWL}, we estimate the
concentration of Kondo magnetic impurities in samples Cu1, Cu2,
Cu3 and Cu4 to be respectively 0.75, 0.3, 1.5 and 1~ppm. Note that
these would be lower bounds on the concentrations if the
paramagnetic impurities are on the surface, with a distribution of
Kondo temperatures.

To conclude, we have measured the Aharonov-Bohm $h/e$ oscillations
to probe the presence of very dilute, low Kondo temperature,
magnetic impurities. By applying this technique to copper wires,
we showed that the saturation of the low-field phase coherence
time is due to spin-flip scattering by magnetic impurities. Recent
measurements of energy exchange in a magnetic field demonstrate
that magnetic impurities are also responsible for the anomalous
energy exchange previously observed in mesoscopic wires
\cite{Anthore}.

We are grateful to I.~Aleiner, A.~Anthore, Y.~Blanter,
M.H.~Devoret, D.~Esteve, V.I.~Fal'ko, L.I.~Glazman, H.~Pothier,
and M.~Vavilov for valuable discussions. This work was supported
by NSF grants DMR-9801841 and 0104178, and by the Keck
Microfabrication Facility supported by NSF DMR-9809688.


\begin{references}
\bibitem{fitWL} See \cite{Gougam}. We used the measured geometrical and electrical
sample characteristics detailed in Table~1. First, we extract the
spin-orbit length $L_{so}$ from the fit at the highest temperature
shown. We found $L_{so}=0.67,$ 0.4, 0.35 and 0.33~$\mu \mathrm{m}$
respectively for samples Cu1, Cu2, Cu3 and Cu4. This leaves
$\tau_\phi(B\approx0)$ as the only fit parameter at lower
temperatures.
\end{references}
\end{document}